\documentclass[aps,superscriptaddress,groupedaddress]{revtex4}  

\usepackage{amsmath, amsthm, amssymb}
\usepackage{graphicx,epsfig}
\usepackage{color}
\usepackage{natbib}
\usepackage{setspace}
\usepackage{fullpage}

\begin{document}
\begin{spacing}{1.25}

\title{Impact Dynamics of Oxidized Liquid Metal Drops}

\affiliation{Department of Physics and James Franck Institute, University of Chicago, Chicago, IL 60637, USA}
\affiliation{School of Natural Sciences, University of California, Merced, CA 95343, USA}

\author{Qin Xu} \affiliation{Department of Physics and James Franck Institute, University of Chicago, Chicago, IL 60637, USA}
\author{Eric Brown} \affiliation{Department of Physics and James Franck Institute, University of Chicago, Chicago, IL 60637, USA}
\affiliation{School of Natural Sciences, University of California, Merced, CA 95343, USA}
\author{Heinrich M. Jaeger} \affiliation{Department of Physics and James Franck Institute, University of Chicago, Chicago, IL 60637, USA}

\date{\today}

\begin{abstract}
With exposure to air, many liquid metals spontaneously generate an oxide layer on their surface. In oscillatory rheological tests, this skin is found to introduce a yield stress that typically dominates the elastic response but can be tuned by exposing the metal to hydrochloric acid solutions of different concentration. We systematically studied the normal impact of eutectic gallium-indium (eGaIn) drops under different oxidation conditions and show how this leads to two different dynamical regimes. At low impact velocity (or low Weber number), eGaIn droplets display strong recoil and rebound from the impacted surface when the oxide layer is removed. In addition, the degree of drop deformation or spreading during the impact is controlled by the oxide skin.  We show that the scaling law known from ordinary liquids for the maximum spreading radius as a function of impact velocity can still be applied to the case of oxidized eGaIn if an effective Weber number $We^{\star}$ is employed that uses an effective surface tension factoring in the yield stress. In contrast, no influence on spreading from different oxidations conditions is observed for high impact velocity. This suggests that the initial kinetic energy is mostly damped by bulk viscous dissipation. Results from both regimes can be collapsed in an impact phase diagram controlled by two variables, the maximum spreading factor $P_m = R_0/R_m$, given by the ratio of initial to maximum drop radius, and the impact number $K = We^{\star}/Re^{4/5}$, which scales with the effective Weber number $We^{\star}$ as well as the Reynolds number $Re$. The data exhibit a transition from capillary to viscous behavior at a critical impact number $K_c \sim  0.1$.
\end{abstract}
\maketitle

\section{Introduction}
Normal impact of liquid metals is important for a wide variety of industrial processes and applications, including electronic fabrication$^{[1]}$, thin film coating$^{[2], [3]}$ and heat conductor production$^{[4]}$. Typically, precise control of the drop spreading is desired, in particular for processes that require reproducible and predictable behavior to build up layers from successive drop impact events$^{[1], [4]}$. However, except for mercury most liquid metals develop an oxide skin when exposed to air$^{[5], [6]}$, resulting in non-Newtonian behavior. One of the direct consequences is that the oxide layer consumes a portion of kinetic energy and thus deviates their spreading dynamics from ordinary fluids in the impact. Also, oxidized liquid metals can generally form non-spherical drops since the surface skin not only prevents contact between air and the bulk of the material but also the minimization of the surface energy. Previous investigations observed the effect of the oxide skin in various measurements, including rheology, wetting capability and surface tension$^{[4],[7]-[9]}$.  However, a detailed experimental study of how the presence of an oxide skin controls the dynamics of liquid metals impinging onto a solid surface has been lacking. This is the focus of the present paper.

In the absence of splashing$^{[10]-[13]}$, previous experiments on Newtonian fluids mainly focused on the geometric deformation of the drop, such as the maximum lateral spreading distance during the impact$^{[14]-[19]}$. Conventionally, this deformation is modeled from an energy conservation point of view, considering the different energy scales affecting the impact and spreading processes: viscous drag dissipates the momentum inside the liquid metal; surface tension stores the initial kinetic energy; and friction from the substrate resists the spreading. We can then associate different regimes with conditions under which certain energy scales dominate the dynamics. For instance, Clanet et al. $^{[16]}$ showed experimentally for Newtonian fluids that the maximum spreading radius scales as a power of the Weber number in the capillary regime and a (different) power of the Reynolds number in the viscous regime. Here, we will show that similar scaling laws can still be used for liquid metals if the relation between skin stress and oxidation condition is appropriately taken care of. 

To realize a systematic control of the skin effect, we use eutectic gallium-indium (eGaIn) as a model fluid, for which oxidation can be managed by immersing the metal in a bath of acid. The acid not only prevents continued oxidation by mitigating the contact with air, but also initiates a chemical reduction process that, in equilibrium, produces a skin thickness corresponding to the particular reaction conditions. Therefore, we can tune the skin by varying the acid concentration and use this to quantitatively analyze the role of the oxide layer in the impact dynamics.
The purpose of the paper is to lay down an experimental framework for describing the impact behavior of liquid metals. The remainder of the paper is organized as follows.   We start with the rheological characterization (Sec. III) to confirm the change of skin-induced stresses under different acid concentrations. Then we qualitatively show how the change of surface elasticity affects the impact behavior on glass (Sec. IV). Within the capillary regime, we demonstrate how unoxidized eGaIn drops can rebound from the substrate at very low Weber numbers (Sec. Va). In order to collapse the spreading data for various surface conditions, we introduce an effective Weber number that depends on drop size and skin stress (Sec. Vb). We then show that the skin effect does not play a significant role in the high-velocity, viscosity-dominated regime (Sec. VI). Finally, we develop a phase diagram that collapses data from both regimes by using appropriately scaled dimensionless groups (Sec. VII), before concluding in Section VIII.

\section{Experimental}
\subsection{Materials}
We purchased the eutectic gallium-indium (eGaIn, purity $99.995\%$) from Sigma-Aldrich. This material has a melting point around $15^0C$. All experiments were performed at room temperature ($\sim 24^0C$ ), thus the materials stayed in its liquid state at all times. At this temperature, liquid eGaIn has a density $\rho=6.25$g/ml  and a nominal viscosity $\eta_l=1.92\times 10^{-3}$Pa$\cdot$ {s}. For the acid bath, we use hydrochloric acid (HCl) of varying concentration. As substrate for the impact experiments, we used a corrosion-resistant glass plate of $8$mm thick.

\subsection{Rheological Measurements}
To characterize the stress response of eGaIn, we performed rheological test in an Anton Paar MCR 301 rheometer with parallel-plate geometry. Figure 1 illustrates our experimental setup. In the measurement, the top rotating tool had a radius $r$ of either $12.7$mm or $4.1$mm. The eGaIn sampls were settled in the gap between rotating top and fixed bottom plate. The gap size $d$  was varied from $1$ to $2$mm. 

For unidirectional shear measurements, the rheometer shears the fluid by rotating the tool at a constant angular rotation rate $\omega$, and the torque $T$ required to shear the fluid is measured in the steady state. The average shear stress is calculated as $\tau=2T/\pi r^3$ and the average shear rate at the plate edge is $\dot{\gamma}=\omega r/d$.

For oscillatory tests, on the other hand, the rheometer tool oscillates back and forth at a fixed frequency and strain amplitude $\gamma$. The storage modulus, defined as $G^{\prime}=\tau/\dot{\gamma}$, provides an indicator of the elastic response to the shear.

To contain both acid bath and sample during the measurements, a cup was constructed consisting of a clear acrylic cylinder with $31.75$mm inside diameter and glued to a titanium plate of thickness $0.51$mm that formed the bottom. Further, the transparency of the cup provides the opportunity to monitor the sample underneath the tool. Due to the high surface energy with most solid surfaces, eGaIn does not stick to the titanium plate.  For this reason, we fill the eGaIn sample to a height larger than the gap size $d$ so there is no extra space for the liquid to spill outward. Also, a $13$mm thick acrylic cylinder was glued to the bottom of the top plate of the rheometer to protect it from the acid (see Fig. 1(a)). From measurements at low shear rate, the stress resolution of our rheometer was found to be $0.06$Pa.  All rheological measurements were performed multiple times to check for repeatability. Because the standard deviation from those repeated tests was $5\%$ or better, in the following we show data from only one measurement  for clarity.

\subsection{Imaging system}
The imaging setup for the drop impact is sketched in Fig.1b. Drops were slowly extruded from a syringe with steel nozzle (inner radius of $0.48$mm and $4.1$mm) by using the RAZEL 99E syringe pump (the pump rate was fixed at 20ml/hr). The falling height was varied from $1$ to $200$cm to give an impact velocity $V_0=(0.4\sim6.3)\pm0.15$m/s to the drop before impinging on the glass substrate. Using an inclined reflective mirror under the substrate, we were able to monitor the entire process from both side and bottom. To highlight the drop profile, the drops were backlit by two white light sources (12V/200W, Dedolight). We used a Vision Research Phantom v12 camera with a macro lens (Nikon Micro $105$mm) to video-capture the impact process. The frame rate reached $17,000$fps at $380\times540$ pixels with a spatial resolution around $15\mu$m. The camera was adjusted to include both the side and bottom images (from the mirror) in the same field of view. Fig. 1c shows typical images captured by the camera.

\begin{figure}[h]
\begin{center}
\includegraphics[width=65mm]{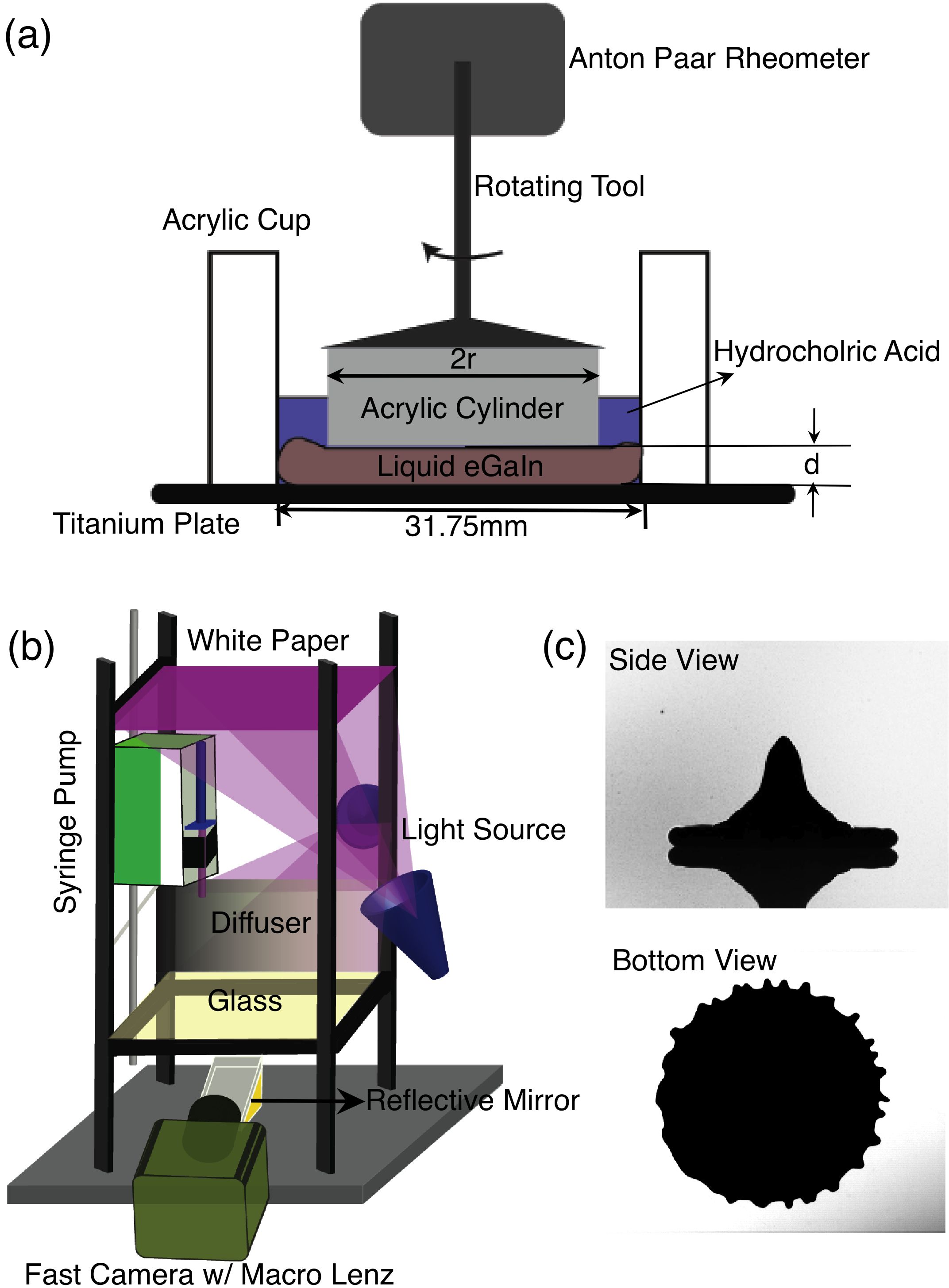}
\end{center}
\caption{(a). Sketch of the rheometer setup. An acrylic cup ($31.75mm$ inner diameter and $13.10mm$ height) was constructed to contain the acid bath, in which the rotating tool shears the eGaIn sample. (b). Schematic representation of our imaging setup. The syringe pump can vertically move along the trail by two meters that is made of four steel rods. Drops during the impact can be lit from the back or the top, depending on the perspective of the video taken by a Phantom v12 camera. The highest frame rate is up to $\sim 10^4$ fps. (c). Two typical images of a liquid eGaIn drop impacting a smooth glass substrate, showing side and bottom views simultaneously captured with the help of the mirror shown in panel (b).}
\end{figure}

For a given nozzle diameter, the initial drop radius   was found to be reproducible with $5\%$ uncertainty or less. To quantify the relative deformation of the drops with time $t$, we use the spreading factor $P$, defined as $P=R(t)/R_0$, where $R(t)$ is the spreading radius obtained by averaging the distance from the advancing contact line to the center of the initial impact and the initial drop radius $R_0$ is measured from the half-width of the horizontal cross-diameter of the drop before impact. All the droplet-related geometric dimensions reported in the following were extracted from the videos using edge-detection algorithms in Matlab.

\subsection{Sample Preparation}
Prior to all rheological measurements, the samples were pre-sheared, or ÒwashedÓ, at $\dot{\gamma}=60$s$^{-1}$ for $10$ mins while immersed in the HCL solution, so that the skin sufficiently reacted with the acid. The eventual chemical equilibrium determines the amount of oxide left on the surface of eGaIn.  The steady state was confirmed by slowly ramping the shear rate first up and then down, and checking there was no hysteresis in the measured stress, except when eGaIn was exposed to air, in which case oxidation is not reversible under shear$^{[7]}$.

Samples used in our study of the impact process were prepared in the same way and also sheared at constant rate of $60$s$^{-1}$ for $10$ minutes. After this washing step, we used a plastic syringe with a steel nozzle tip to extract eGaIn from the bath. Then the syringe was inserted into the pump and liquid eGaIn was slowly ejected from the nozzle to form the drops. Since the volume of a single droplet was around $0.15$ml, it took approximately $0.15$ml$/(20$ml$/$hr$) = 20$s to fully extrude the drop until it fell off. Together with the time spent in mounting the syringe, impact occurred in less than $30$s after the washing step. Also, to avoid possible oxidation during the time of performing multiple impact experiments, the data we used is usually from the first drop extruded from the nozzle. Therefore, fresh oxidation, which usually initiates after a sample has been exposed to air for a couple of minutes, can be neglected for our measurements.

\section{Rheology Results}

\begin{figure}[h]
\begin{center}
\includegraphics[width=85mm]{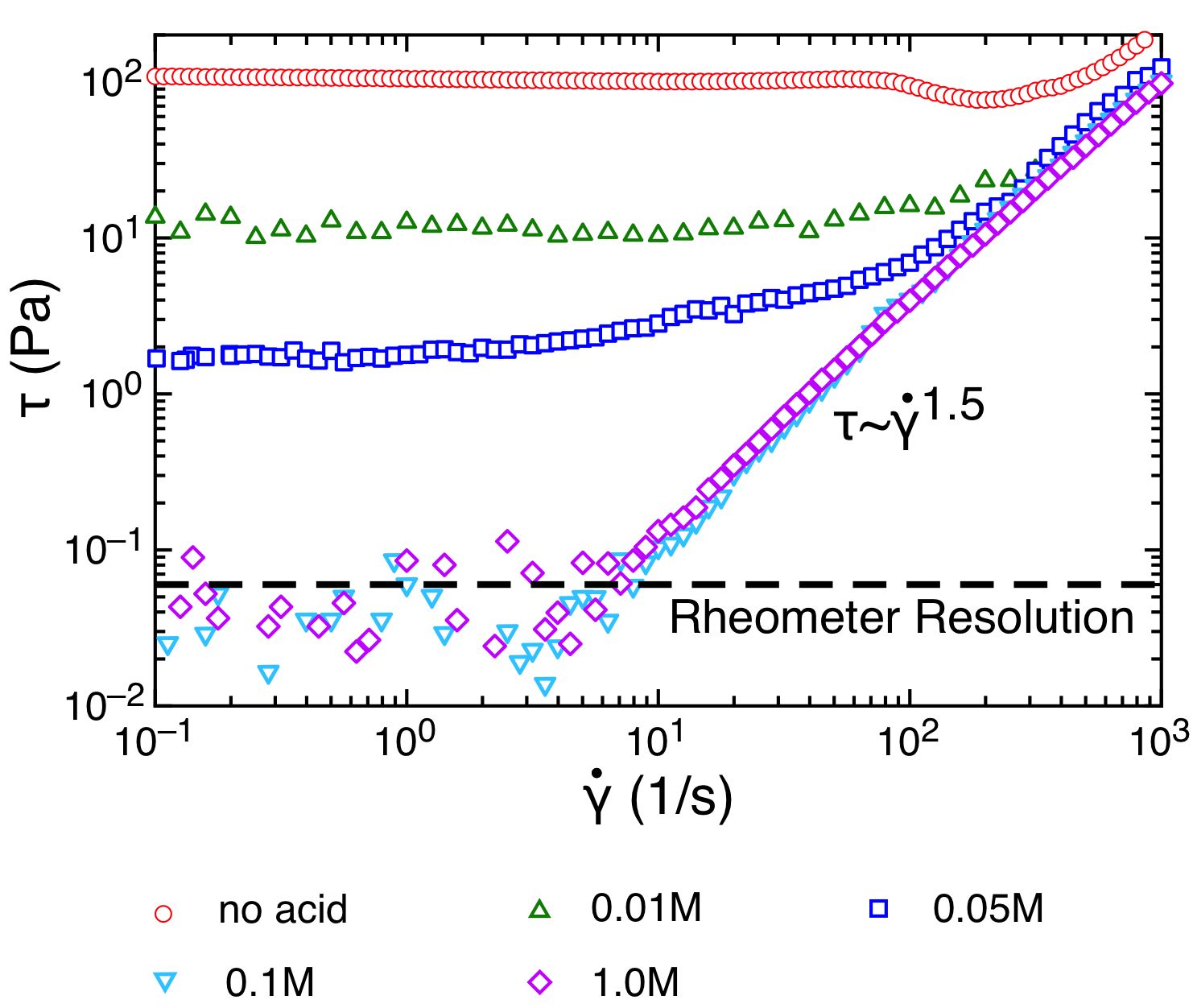}
\end{center}
\caption{Shear stress vs. shear rate curves for liquid eGaIn in HCl baths with different acid concentrations.A detailed discussion of the similar result  can be referred to [7] (Sec. III. A. 1. , Fig. 4).}
\end{figure}

The drop impact dynamics of non-Newtonian fluids are usually modeled based on their intrinsic rheological properties. However, we have shown previously that surface oxidation is crucial for the shear stress measurement of eGaIn in the steady state$^{[7]}$: for instance, Fig. 2 provides plots of shear stress $\tau$ against shear rate $\dot{\gamma}$ for eGaIn submerged in acid baths of different concentrations. Initially, $2$ml of eGaIn was directly placed on the bottom plate of the rheometer and exposed to air. Such an oxidized droplet appears dirty and wrinkled on the surface, while the bulk appears shiny and mirror-like if we slice open the skin. When exposed to air (red upward pointing triangle points in Fig. 2), eGaIn displays a significant yield stress $\tau_y\sim10^2$Pa, indicating an effective solidification of the material.  Adding HCl into the bath can eliminate the skin effect. As shown in Fig. 2, when the acid concentration reaches $C_{HCl}>0.1$M, $\tau_y$ dramatically drops by four orders of magnitude to nearly zero within the rheometer resolution. 

This vanishing yield stress at sufficiently high acid concentration suggests that pure eGaIn behaves as an ordinary viscous fluid, for which the dynamic viscosity is given by the ratio between shear stress and shear rate in the laminar regime. However, because of the large density of eGaIn, inertial effects can easily play a more essential role in the shear flow than they would for normal fluids. As a result, at high shear rate all traces in Fig. 2 collapse onto the typical scaling of inertial flow$^{[7], [20]}$, $\tau\sim\dot{\gamma}^{3/2}$ (solid line).

Fig.2 confirms that the measured yield stress of the eGaIn samples in air is associated with oxidation. The relative contributions to the overall stress from the bulk of the material and from the skin can be extracted from measurements on samples of different sizes. To this end, we did oscillatory viscoelasticity measurements using two different sizes for the rotation plate, $8.2$mm and $25.4$mm in diameter, with correspondingly different sample volumes and exposed surfaces around the perimeter of the plate.

Performing strain oscillations at fixed angular frequency $\omega=0.5$ rad/s and gap size $d=1$mm, the applied average strain $\gamma$ was ramped from $0.001$ to $10$. For the oxidized eGaIn sample, over the range of $\gamma<0.1$, the elastic modulus $G^{\prime}$ of the oxidized eGaIn was found to be two orders of magnitude larger than the loss modulus  $G^{\prime\prime}$ (this is similar to the observation by Larsen et al.$^{[5]}$, except that they used $\omega=1$ rad/s).

For ordinary viscoelastic materials, whose stress originates from the elastic response of the bulk, $G^{\prime}$ is supposed to be an intrinsic quantity, independent of sample size. However, the apparent elastic modulus of oxidized eGaIn is found to be more complicated. Fig.3a shows $G^{\prime}$  against strain $\gamma$ at different acid concentrations. When the sample is exposed to air (red circles) or $0.01$M acid solution (blue circles), the plot shows significant size dependence: $G^{\prime}$ is reduced by a factor around three when the plate diameter is increased by the same factor. In addition, a decrease of  $G^{\prime}$ starts to occur at  $\gamma\sim10^{-2}$, which corresponds to the end of the linear elastic regime and signals breaking of the skin. By contrast, for $C_{HCl}>0.1$M  (pink and green circles), the obvious size dependence of $G^{\prime}$ disappears. Instead, over the range of $\gamma$ scanned, $G^{\prime}$ constantly stays below the rheometer resolution limit ($\sim 0.06$Pa, the dashed black line in the figure), which indicates a very small bulk modulus of liquid eGaIn. Thus, the behavior of $G^{\prime}$ more consistent with ordinary fluids is recovered at high acid concentration, or weak skin effect. In other words, $G^{\prime}$ may not be an appropriate, intrinsic parameter when the skin exists.

\begin{figure}[h]
\begin{center}
\includegraphics[width=85mm]{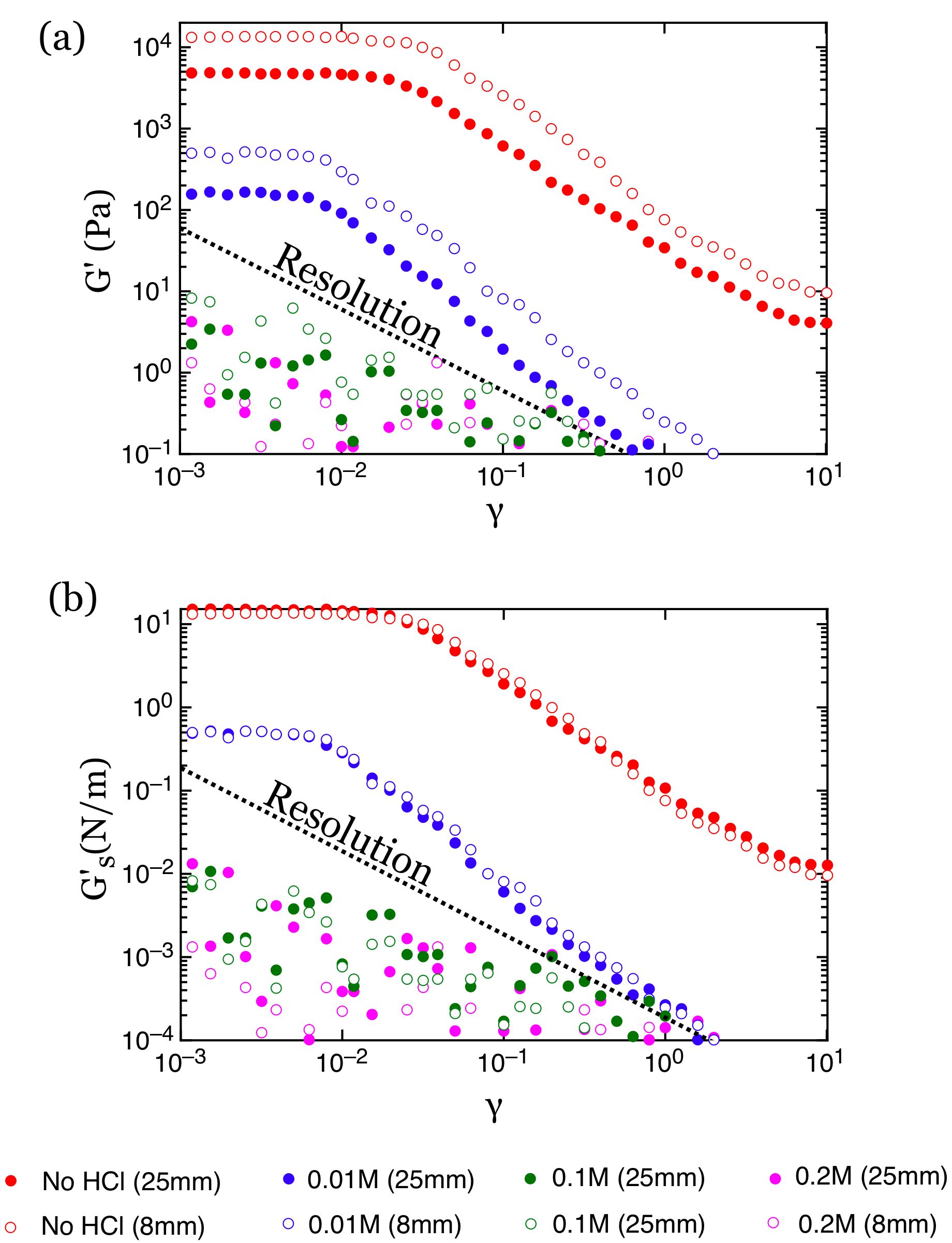}
\end{center}
\caption{Apparent bulk (a) and surface (b) elastic moduli as a function of applied average strain for eGaIn under $\omega=0.5$Hz oscillatory tests.  Different colors (red, blue, pink, green) indicate experiments performed at various acid concentrations. Plates with two diameters, $25$mm (solid circle) and 8mm (open circle), were used in the measurements. The dotted straight lines represent the rheometer resolution. }
\end{figure}

We can model the sample size dependence by considering the elastic response of the sample as originating from two sources, the bulk and the skin, and treating the skin as a very thin, effectively two-dimensional layer. Inside such layer, in direct analogy to a bulk stress, the surface stress $\tau_s$ relates the in-plane surface strain $\gamma_s$ to the surface elastic modulus $G^{\prime}_s$ by
\begin{equation}
\tau_{s}=G^{\prime}_s \cdot \gamma_s
\end{equation}

\noindent Here, $\tau_s$ and $G^{\prime}_s$ are defined as forces per unit length, i.e., represent a surface tension, and are intrinsic quantities characterizing the surface. Generally, the total measured torque $T$ incorporates contributions from both the skin and the bulk of the material, so that
\begin{equation}
T=2\pi r^2 (G^{\prime}_s \gamma)+\frac{1}{2}\pi r^3 (G^{\prime}_b\gamma)
\end{equation}
$G^{\prime}_b$ is the regular bulk modulus. Also, we assume that there is no relative displacement between the skin and bulk so that  $\gamma_s=\gamma$. Using $\tau=2T/\pi r^3$ and defining the measured, effective elastic modulus of the sample as  $G^{\prime}=\tau/\gamma$, we have
\begin{equation}
G^{\prime}=4G^{\prime}_s/r+G^{\prime}_b
\end{equation}
Therefore, when the surface elasticity dominates the shear response ($G^{\prime}_s/r\gg G^{\prime}_b$), the measured modulus  $G^{\prime}$ can be directly related to the surface modulus by $G^{\prime}_s=(G^{\prime}\cdot r)/4$.$^{[5]}$ 

Fig. 3b shows the same data as Fig. 3a, but rescaled by multiplying with ($r/4$). Now the traces for different plate diameters collapse at low acid concentrations (zero or $0.01M$ HCl), while they are still kept below the resolution (dashed line) at high acid concentrations ($C_{HCl}$). This indicates that indeed the surface oxidation dominates at low acid concentration and that the observed size dependence of the measured modulus $G^{\prime}$ in Fig. 3a reflects the properties of the skin. In addition, since the data is reversible in different acid concentrations, the skin effect is localized on the surface and does not affect the properties of the bulk.  Together, Figs. 3a and b demonstrate that the tiny bulk stress $G^{\prime}_b$ only shows up when the oxidation is removed. 

Therefore, for sufficiently oxidized eGaIn $G^{\prime}_s$ should be used as intrinsic parameter, while in more concentrated acid baths the measured $G^{\prime}$ is the appropriate intrinsic parameter, reflecting the bulk modulus of pure eGaIn. 

\section{Drop Impact under Different Oxidation}

\begin{figure}[h]
\begin{center}
\includegraphics[width=85mm]{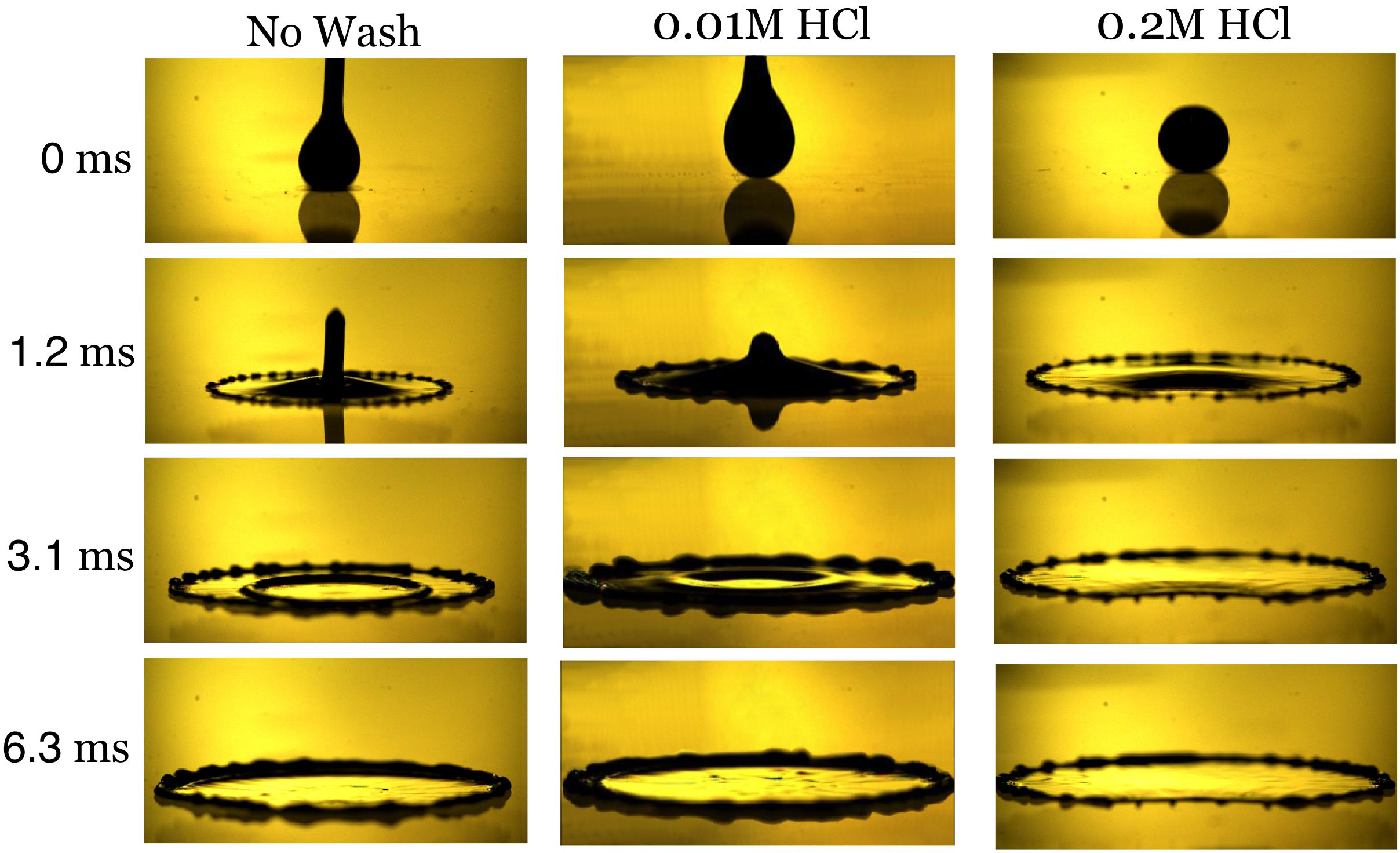}
\end{center}
\caption{Typical image sequences of eGaIn drops impacting onto a glass substrate. Drops are initially washed in HCl solution as indicated. For all three images sequences shown above, the impact velocity was kept at  $V_0=(1.02\pm0.12)$m/s and the initial drop diameter was  $R_0=(6.25\pm0.10)$mm.  }
\end{figure}

By ejecting eGaIn from the same nozzle and the same falling height, we generated droplets with reproducible impact velocity and radius, $V_0=(1.02\pm0.12)$m/s  and  $R_0=(6.25\pm0.10)$mm. Fig. 4 shows three image sequences for eGaIn drops with different skin strengths. The left column shows the impact of an air-oxidized eGaIn drop not pre-washed in acid. A long tail at the top end of the drop is formed when the fluid detaches from the nozzle. Different from ordinary liquids, the oxide skin prevents the fluid to freely relax the surface energy, so that this non-spherical geometry is kept unchanged during the falling stage. After the impact occurred, a thin liquid metal sheet or ÒlamellaÓ expands rapidly along the smooth substrate. 

At $t=1.2$ms, the lower end of the tail structure reaches the thin liquid sheet, which stimulates a secondary impact at the liquid-liquid interface. The induced surface wave (Fig. 4, the third image in the left column) propagates along the liquid sheet with the speed of shallow water waves$^{[21]}$ and eventually catches up with the liquid-glass contact line (at $6.2$ms). The surface wave is generally found in all impacts of oxidized eGaIn because of the non-spherical shapes of the initial drop.
 
In contrast, cleaning the samples with acid removes the oxide and weakens the skin effect. The middle and right columns in Fig. 4 show images of drops pre-washed in $0.01$M and $0.2$M HCl respectively. Since the yield stress is reduced to $\sim10$Pa at $0.01$M HCl, the length of the tail becomes much shorter (middle column) and only a very week surface wave appears ($t=3.1$ms). If we keep increasing the HCl concentration to $0.2$M, the acid becomes strong enough to fully eliminate any observable skin effect, and eGaIn shows no difference in the spreading behavior with ordinary liquids (right column). Generally, we do not observe any splashing of eGaIn since the surface tension ($>400$mN/m) is much larger than ordinary liquids. 
 
 Another major feature exhibited in Fig. 4 for a relatively large drop ($R_0\sim 6$mm) is the significant variance in the maximum spreading radius with and without oxide skin. Simple inspection of the images gives a $20\%$ difference in the final stage radius between left and right column, a result that is repeatable within $<5\%$ experimental uncertainty. Therefore, the skin not only affects the drop shape but also resists the spreading of the contact line along the glass substrate. Since the only controlling parameter is the acid concentration, the surface elasticity must play a critical role in determining the maximum spreading distance.  
 
 \section{Capillary Regime}
 For any drop impacting on a hard surface, its kinetic energy immediately before impact will play a key role in the subsequent outcome. When the impact speed is low and the surface is relatively smooth, the drop typically spreads out and comes to rest as a truncated sphere or thin, disc-like film, depending on the equilibrium contact angle$^{[22]}$. When the impact velocity is sufficiently small to neglect viscous dissipation during droplet deformation, the balance between kinetic and surface (or ÒcapillaryÓ) energies becomes the controlling factor. In this capillary regime, the dimensionless Weber number, $We$, gives the ratio of inertial to surface stresses. Conventionally, for a drop of radius $R_0$, surface tension of $\sigma$ , density of $\rho$ and impact velocity of $V_0$ , the Weber number is defined as
 \begin{equation}
 We=\frac{2\rho V_0^2 R_0}{\sigma}
 \end{equation}
 \noindent In our experiments, due to the dramatically high surface tension of liquid eGaIn, the capillary regime is easy to achieve and commonly observed even at moderately large impact velocities.
 
\subsection{Rebound Control at Low Weber Number}
\begin{figure}[h]
\begin{center}
\includegraphics[width=90mm]{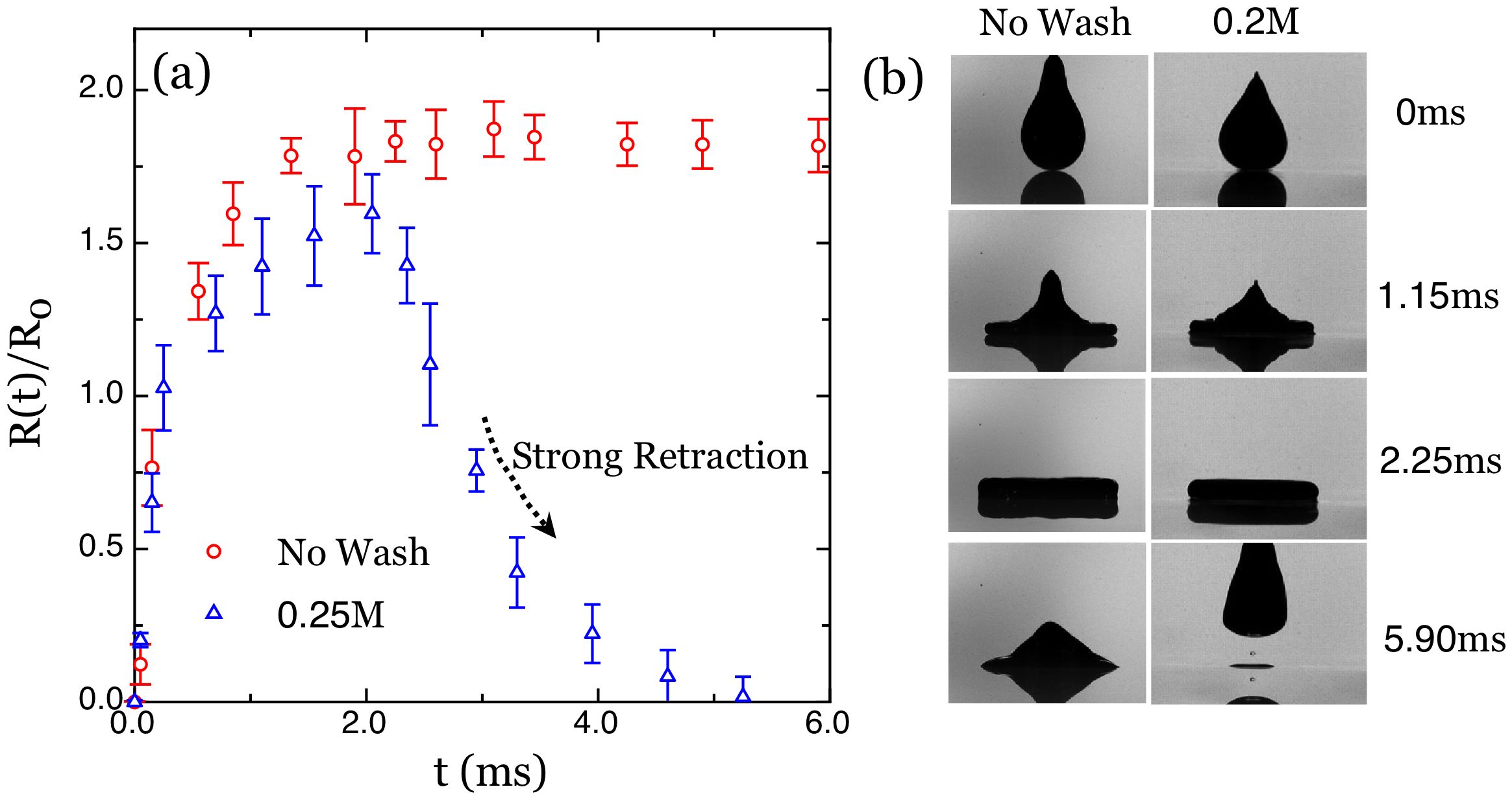}
\end{center}
\caption{(a). Relative spreading radius $R(t)/R_0$ as a function of time after impact for both unwashed (red) and HCl-washed (blue) eGaIn drops impacting a glass surface at velocity $V_0=(0.83\pm0.13)$m/s. The experiments were repeated five times; the error bars correspond to the standard deviation (b). Image sequences for comparing the impact of a skin-covered drop (left), which sticks to the glass, with that of a non-oxidized drop (right), which rebounds from the surface. }
\end{figure}

One of the striking outcomes that can be observed in the capillary regime is drop rebound. This is usually achieved by coating the substrate with a thin layer of super-hydrophobic material (such as wax or polymer)$^{[16], [23]}$, in order to form a large contact angle ($>160^0$ ) between drop and substrate. In this situation, kinetic energy is rapidly converted into surface energy at the drop/air interface during the spreading. After the spreading front reached its maximum radius, the built-up surface energy can be released by pulling the contact line backward and potentially lifting the drop off the surface. 

Previous studies have shown that the skin effect determines the degree to which eGaIn wets glass$^{[4,7]}$. Generally, oxidized eGaIn wets most solid surfaces well, while pure, unoxidized eGaIn becomes perfectly non-wetting $(\sim 180^0)$. Intermediate contact angles can be set by the acid concentration used to wash the sample. Therefore, a precise rebound control of eGaIn drops can be carried out by adjusting the acid bath.

Fig. 5 compares the behavior of drops of the same size but using oxidized and pure (washed in $0.2$M HCl) eGaIn as they impact a glass surface with equal impact velocity $V_0=(0.83 \pm 0.13)$m/s.  At each time step, the spreading radius $R(t)$ was measured and $R(t)/R_0$ is plotted against time (Fig.5a). For oxidized eGaIn (red upward pointing triangles), the drop expands rapidly along the surface until the spreading front arrives at the maximum   $R_m\sim1.8 R_0$ at $t~2$ms. Subsequently, even though the fluid slightly varies its surface shape to relax the interfacial energy, the contact line stops moving forward. Therefore, $R(t)$ keeps its value at $\sim1.8R_0$. 

The behavior of a drop of eGaIn washed in $0.2$M HCl at least initially ($t<1$ms) appears similar to that of the skin-covered drops, since spreading during this early stage is mostly due to geometric deformation. The maximum spreading radius, $R_m \sim 1.7 R_0$ in this case, is also reached at similar time ($t\sim$2ms). However, instead of coming to rest, a strong retraction of the fluid is seen immediately after reaching $R_m$, and  eventually drops to back to zero at $t=5.25$ms, which corresponds to the moment when the drop completely detaches from the substrate (see Fig. 5b). 

Drop rebound requires that a large portion of surface energy be converted back into kinetic energy during the retraction stage. Thus, even though the system is still in the capillary regime, a small amount of viscous dissipation or friction can cause the retraction speed to decay, preempting detachment. We therefore expect there to be an upper limit for the spreading radius or, equivalently, for the impact velocity, above which dissipation becomes large enough to eliminate rebound. 

To test this, we carried out a series of experiments with drops of $0.2$M HCl washed, unoxidized eGaIn, parameterizing the impacts by the Weber number. Fig. 6 shows traces of $R(t)/R_0$ with fixed $R_0\sim0.98$mm as function of time for different initial Weber numbers ($We=5.6, 19.8, 22.6, 43.2$). Since all samples were sufficiently acid-washed, the skin effect does not need to be considered. 

\begin{figure}[h]
\begin{center}
\includegraphics[width=85mm]{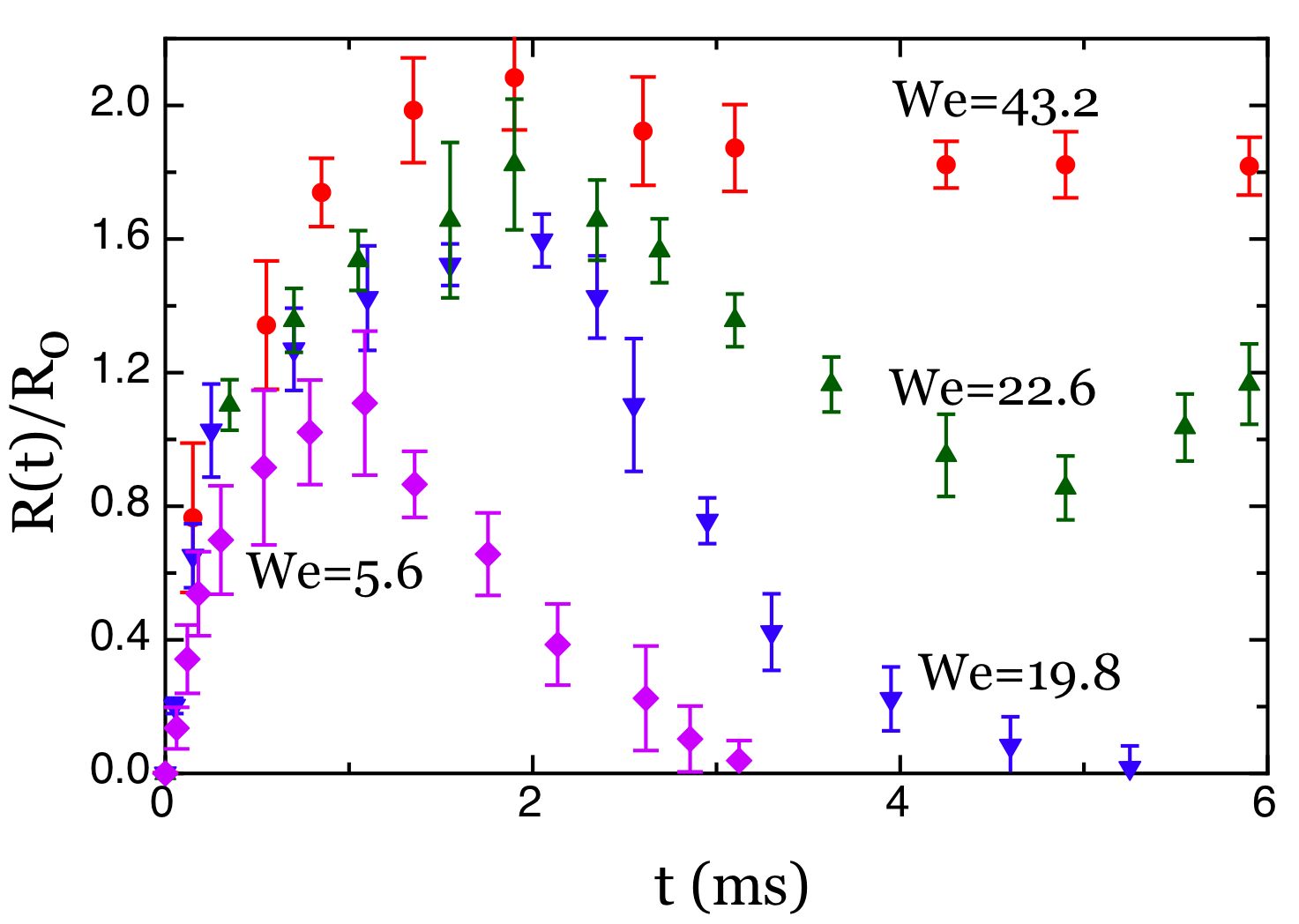}
\end{center}
\caption{Weber number dependence of $R(t)/R_0$ for liquid eGaIn washed by $0.2M$ HCl. The initial drop was fixed at $R_0 = 0.98$mm. The impact velocity is then chosen to be $V_0=0.45$m/s, $0.83$ms, $0.90$m/s and $1.25$m/s. As a result, the corresponding Weber numbers are $We = 43.2, 22.6, 19.8$ and $5.6$.}
\end{figure}

Clearly, the rebound has disappeared for  $We=22.6$ and $43.2$. Instead, without losing the contact of the substrate, a weak retraction follows the maximum radius ($t\sim1.9$ms). After the receding velocity vanishes ($t\sim4.7$ms), the remaining kinetic energy causes the inertial oscillation of the contact line. Finally, the radius stops moving when all the kinetic energy is lost from viscous dissipation. On the other hand, at relatively low Weber numbers,  $We=5.60$ and $19.8$, the detachment between the drop and surface shows up as expected. Therefore, the critical Weber number $We_c$, indicating the upper limit of the rebound regime, is around 20. 

In fact, there should be also a lower Weber number limit to observe rebound, since kinetic energy may not be large enough to lift up the drop if the impact velocity is too small. However, in order to explore this regime, the drop height has to be so small ($\ll1$cm) that residual fluctuations in drop velocity as well as drop size, unavoidable with our set-up, give rise to large uncertainties. Thus, we were not able to resolve any lower bound from our experiments.

\subsection{Scaling of the Deformation}
The discussion so far proved qualitatively that the skin can alter the impact behavior in the capillary regime. In this section, a more quantitative description of the role of the surface elasticity is used as a one step further toward developing a scaling of the maximum spreading factor $P_m=R_m/R_0$ with impact parameters. For Newtonian fluids, $P_m$ is known to scale as $We^{1/4}$, reflecting a momentum balance between inertial and surface surface tension$^{[16]}$. The universality of this scaling has been tested across a wide range of materials, including water, alcohol, viscous glycerin and liquid mercury, allowed for large changes in the intrinsic parameters such as density, viscosity and surface tension$^{[16]}$.  Nevertheless, none of these experiments involved a situation where a portion of surface energy is elastically stored in a surface skin.  

In our experiments we controlled the impact by varying velocity $V_0$, nozzle radius $R$ and oxidation degree. The impact velocity was kept under $2m/s$. Therefore, the ratio of viscous to surface energy $E_{\mu}/E_{\gamma}=3CaP_m^4/8<0.15^{[24]}$, so that viscous dissipation was much less important in this regime (here $Ca=\mu v/\gamma$ is the Capillary number). Meanwhile, we used nozzles with two radii, $R=0.48$mm and $4.1$mm, to test for size-dependence.  

Fig. 7a plots  $P_m$ vs. $We$ under these conditions. Different colors stand for different surface oxidation levels. The solid symbols indicate data taken with the small nozzle while the open symbols correspond to the large nozzle. For the eGaIn sample washed with $0.2$M HCl before impact to remove any skin (red data points), the data for both nozzle sizes collapses onto the conventional $We^{1/4}$ scaling (dashed line). This confirms that pure eGaIn follows the behavior of Newtonian fluids, for which the scaling of $P_m$ with $We$ should be independent of the drop size. Similar results are also observed for drops washed with $0.01$M HCl (pink), except that the entire data set shifts downward by about $5\%$. 

However, the spreading of oxidized eGaIn exhibits a significant dependence on the size of the nozzle. For a given Weber number $We$, $P_m$ for the large nozzle (open blue symbols) is seen to fall below that for the small nozzle (solid blue symbols), shifting downward by approximately $60\%$. This is inconsistent with the properties expected of Newtonian fluids. We note that for the oxidized eGaIn drops from the $0.48$mm nozzle,  $R_m/R_0$ remains consistent with the conventional scaling $P_m\sim We^{1/4}$ and significant variance of $P_m$ occurs only for the $4.1 mm$ nozzle. This suggests that the skin effect becomes important only when the drop size gets sufficiently large. 

\begin{figure}[h]
\begin{center}
\includegraphics[width=90mm]{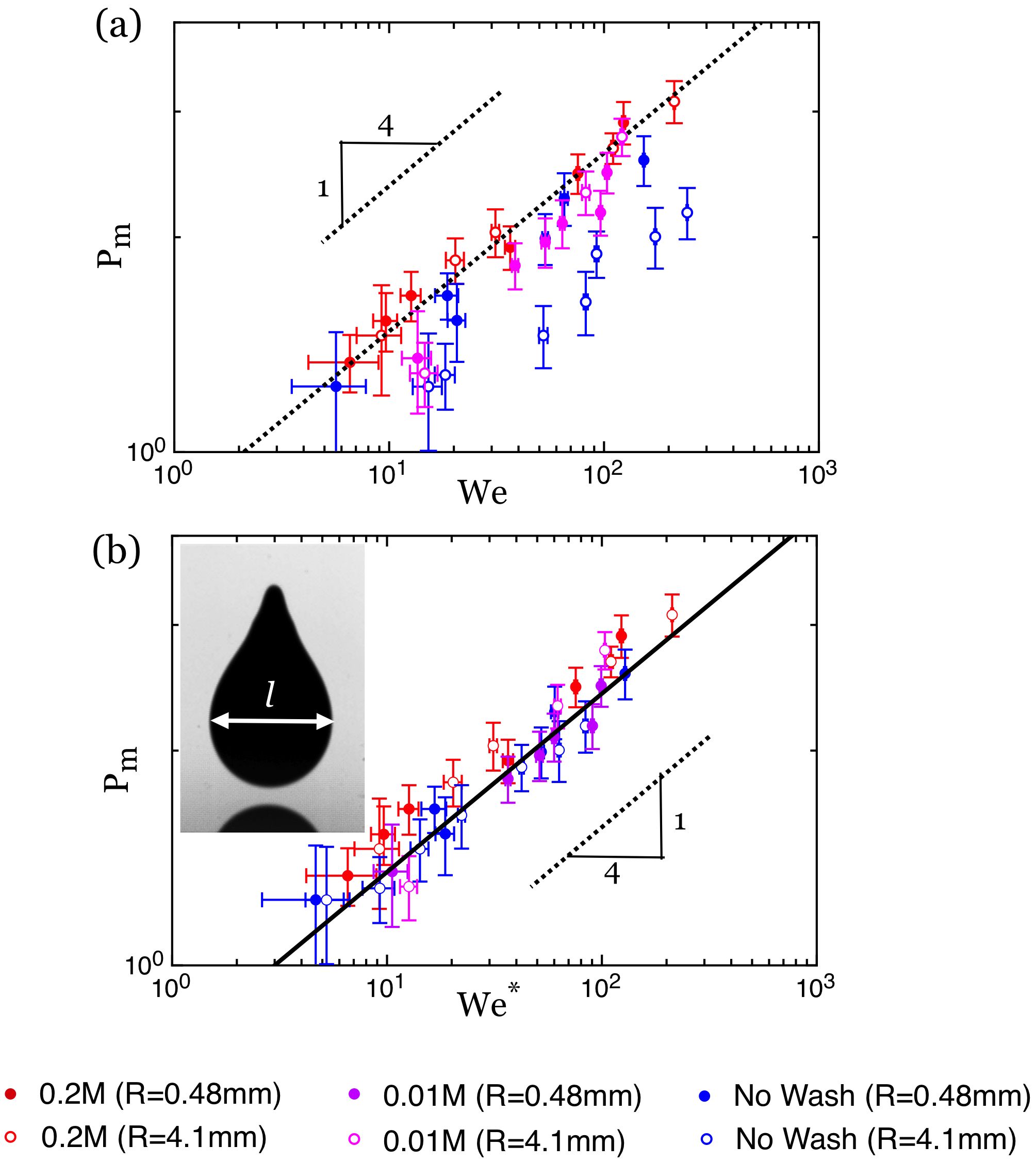}
\end{center}
\caption{(a). Spreading factor $P_m$  vs. Weber number $We$ (a) and effective Weber number $We^{\star}$(b) under different experimental conditions. Colors (red, blue, pink, green) indicate experiments performed at different oxidation conditions. Data for two nozzle radii, $R=0.48$mm (solid circles) and $R=4.1$mm (open circles), are shown. The inset of (b) shows the diameter $l(=2R_0)$ of an eGaIn drop.}
\end{figure}

Based on the comparisons between the drop mass and $R_0$, the uncertainty introduced by non-spherical shape is less than 8\%, which is not large enough to give rise to the scatter of data in Fig. 7 (a). Instead, This drop size dependence of $P_m$ can be explained by the rheology results discussed in Sec. III. If the sample is oxidized, the skin induces extra in-plane stress ($\tau_s$) in the skin surface. As a consequence, the effective surface tension includes contributions from both the bulk ($\sigma$) and the skin ($\tau_s$) and can be written as $\sigma_{eff}={\sigma+\tau_s}=\sigma+G^{\prime}_s\gamma_s$. As discussed in Sec. III, $G^{\prime}_s$ is connected to bulk modulus $G^{\prime}$ by multiplying with the size scale of the eGaIn sample $l_c$(which is the tool radius for the rheology measurement), $G^{\prime}_s=(G^{\prime}\cdot l_c)/4$. Thus,
\begin{align}
\notag \sigma_{eff} &=\sigma+\tau_s=\sigma+G^{\prime}_s \gamma_s \\
&=\sigma+\frac{1}{4}(G^{\prime}\cdot l_c)\gamma_s=\sigma+\frac{1}{4}(G^{\prime}\cdot\gamma_s)l_c
\end{align}
\noindent Since there is no relative displacement between the bulk and skin ($\gamma=\gamma_s$) and the measured stress of oxidized eGaIn directly gives the yield stress $\tau_y$, ${G^{\prime}}\cdot\gamma_s=G^{\prime}\cdot \gamma=\tau_y$. Accordingly, equation (5) can be expressed as  $\sigma_{eff}=\sigma+\tau_y l_c/4$. Previous experiment$^{[7]}$ on the surface tension of different sizes eGaIn drops quantitatively proved that the size scale $l_c$ of pendant drop is simply the diameter $l=2R_0$. As a result, the effective surface tension scales as
\begin{equation}
\sigma_{eff}=\sigma+\frac{1}{2}\tau_y R_0.
\end{equation}
This model fits the apparent surface tension ($\sigma$) from the pendant drop sizes$^{[7]}$.

Physically, the size dependence of $\sigma_{eff}$ comes from the increase of $\tau_s$ with the drop size. However, both $\tau_s$ and $\gamma_s$ are difficult to directly measure in the experiment.. By relating the surface parameters ($G^{\prime}_s, \gamma_s$) with the measured bulk modulus ($G^{\prime}$) and yield stress ($\tau_y$) in Eq. (5), we are be able to quantitatively calculate $\sigma_{eff}$ via Eq. (6).

If we now account for the surface energy stored in the skin by replacing  $\sigma$ in $We$ with $\sigma_{eff}$ given by Eq. (6) we can define an effective Weber number as
\begin{equation}
We^{\star}=\frac{2\rho V_0^2R_0}{\sigma_{eff}}=\frac{We}{1+\tau_yR_0/2\sigma}
\end{equation}
The native surface tension of pure liquid eGaIn is $\sigma\sim4\times10^2 $mN/m. Pre-washing the sample with $0.2$M HCl reduces the yield stress to $0.1$Pa (Fig.2). Since the largest drop made in the experiment has the cross diameter $R_0\simeq 6$mm,  $\tau_y R_0/(2\sigma)<0.01\ll1$ and hence $We^{\star}\simeq We$. Eq. (7) also explains why the data for small, oxidized eGaIn drops follows the conventional scaling even though the skin is present. However, when the product of  $l$ and $\tau_y$ becomes large, $We^{\star}$ can differ significantly from $We$. For instance, with yield stress in air $\tau_y\sim10^2$Pa  for an oxidized drop of size $R_0\sim 5$mm,   $\tau_y R_0/2 \sigma\sim 1$ and $We^{\star}$ drops below $We$ (open blue symbols in Fig. 7a). To test the validity of Eq. (7), we replot $P_m$ from Fig. 7a against the rescaled Weber number $We^{\star}$ in panel (b). Within the experimental uncertainty, the data collapse nicely onto $P_m\sim We^{\star 1/4}$. 

	Since the impact velocity was below $2$m/s in these experiments, the impact shear stress was below the order of $10^2$Pa. In this regime the loss modulus of the skin is smaller than the elastic modulus by two orders of magnitude as mentioned earlier. Hence, viscous dissipation in the skin during impact is negligible and this is why the role of the skin can be represented simply by an effective Weber number.

\section{viscous regime}
\begin{figure}[h]
\begin{center}
\includegraphics[width=80mm]{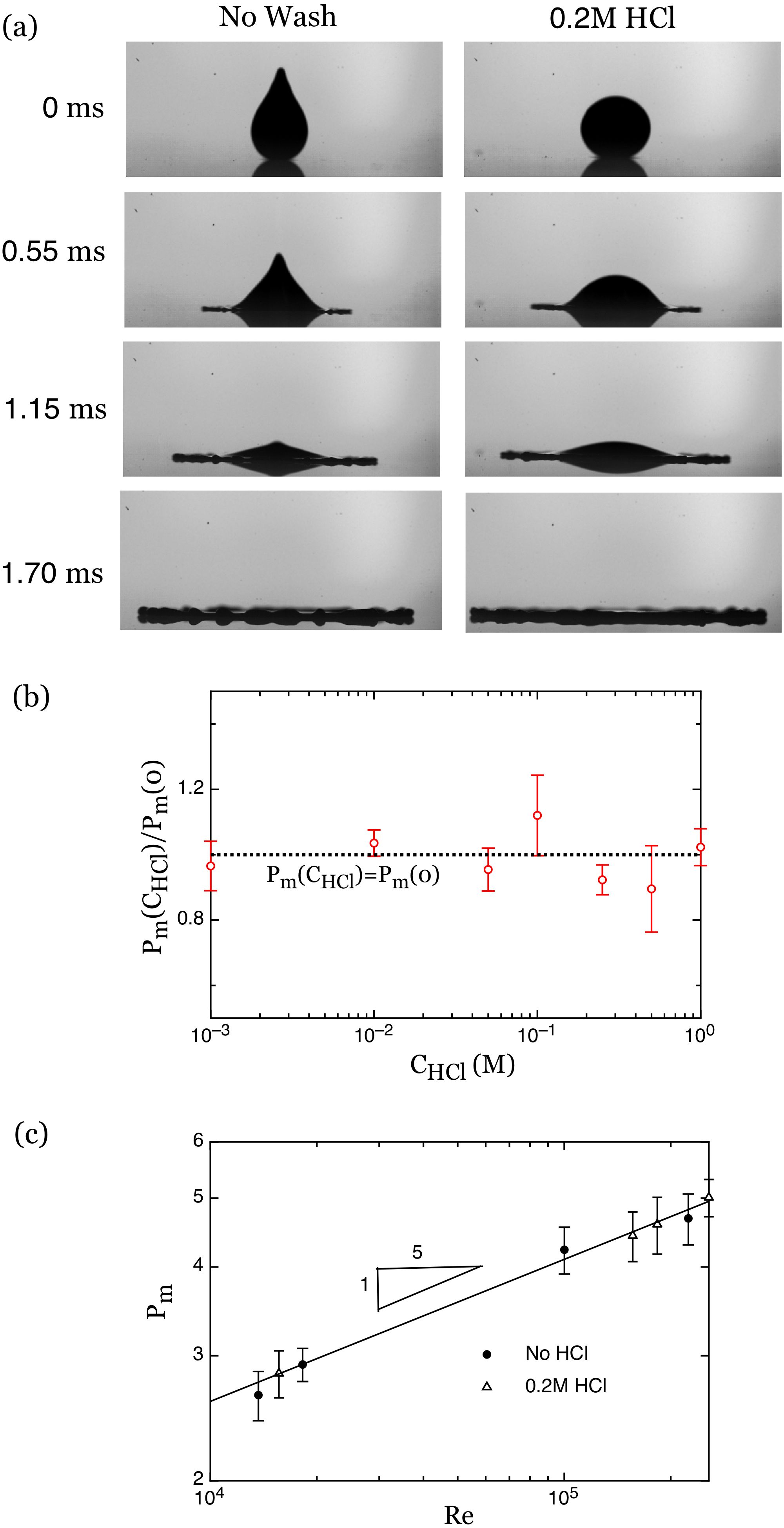}
\end{center}
\caption{(a). Image sequence of impact process of liquid eGaIn in the viscous regime. In particular, the images on the left and right column are taken from oxidized and pure eGaIn respectively. (b). the ratio of $P_m(C_{HCl})/P_m(0)$ is plotted against the acid concentrations of acid solution used to pre-wash the sample. (c). The classic one-fifth power law can be applied to the data of $P_m$ vs. $Re$ in the viscous regime. The Reynolds number is varied by changing the impact velocity and nozzle size ($4.1$mm and $0.48$mm).}
\end{figure}

The surface tension of liquid eGaIn is about one order of magnitude larger than in normal fluids. Therefore, the liquid/gas interfacial energy is more important than viscous dissipation in most impact experiments. It is also the reason why most previous experimental data for liquid metals$^{[14,] [18], [25]}$ was restricted to the capillary regime. 

In order to observe the role of viscous damping, the impact velocity has to be increased such that a large part of initial energy is dissipated through viscous resistance. Experimentally, faster impact was achieved by mounting the syringe pump at the top of a long steel rail to increase the drop release height to as large as two meters.  This extended the impact velocity range up to $6.3$m/s.

In Fig. 8 we compare results for eGaIn drops from the large nozzle ($R=4.8$mm), one air oxidized and another pre-washed with $0.2$M HCl. The diameter of the drops was  $(12.3\pm0.5)$mm. Fig. 8a displays the image sequences during the impinging of the two drops at impact velocity $3.8$m/s. Despite the difference in the initial drop shape, the maximum spreading diameter $P_m$ approximately stays the same.  This independence of $P_m$ is also found when altering the surface oxidation conditions. In Fig. 8b, $P_m(0)$ indicates the maximum spreading factor of unwashed eGaIn drops while $P_m(C_{HCl})$ represents that of drops washed in an acid bath of specified concentration. Varying $C_{HCl}$ over three orders of magnitude, from $0.001$M to $1.0$M, the ratio of $P_m(C_{HCl})$  to $P_m(0)$ is seen to fluctuate around unity within the experimental uncertainty.  This demonstrates that the skin effect is not important in this regime and most viscous dissipation is caused in the bulk rather than by the surface.

 For Newtonian fluids, the spreading factor scales as  ${P_m\sim Re^{1/5}}^{[16]}$ in the viscous regime. Here the Reynolds number indicates the ratio of inertial to viscous stresses and is defined as
 \begin{equation}
 Re=\frac{2\rho V R_0}{\mu}
 \end{equation}
\noindent Given that the skin hardly affects the behavior in the viscous regime, we expect this scaling to survive even for oxidized drops. Indeed, by applying a power law fit to the data within more than one decade of $Re$ and different oxidation conditions (Fig. 8(c)), we obtain  $P_m\simeq0.41Re^{1/5}$. It is worth mentioning that, even though the Reynolds number here is around the order of $10^4$, we do not consider the dissipation caused by turbulence since the time scale of impact ($\sim$ a couple of milliseconds) is too short to form any inertial flow in the fluid.

\section{Capillary to Viscous Transition}
\begin{figure}[h]
\begin{center}
\includegraphics[width=80mm]{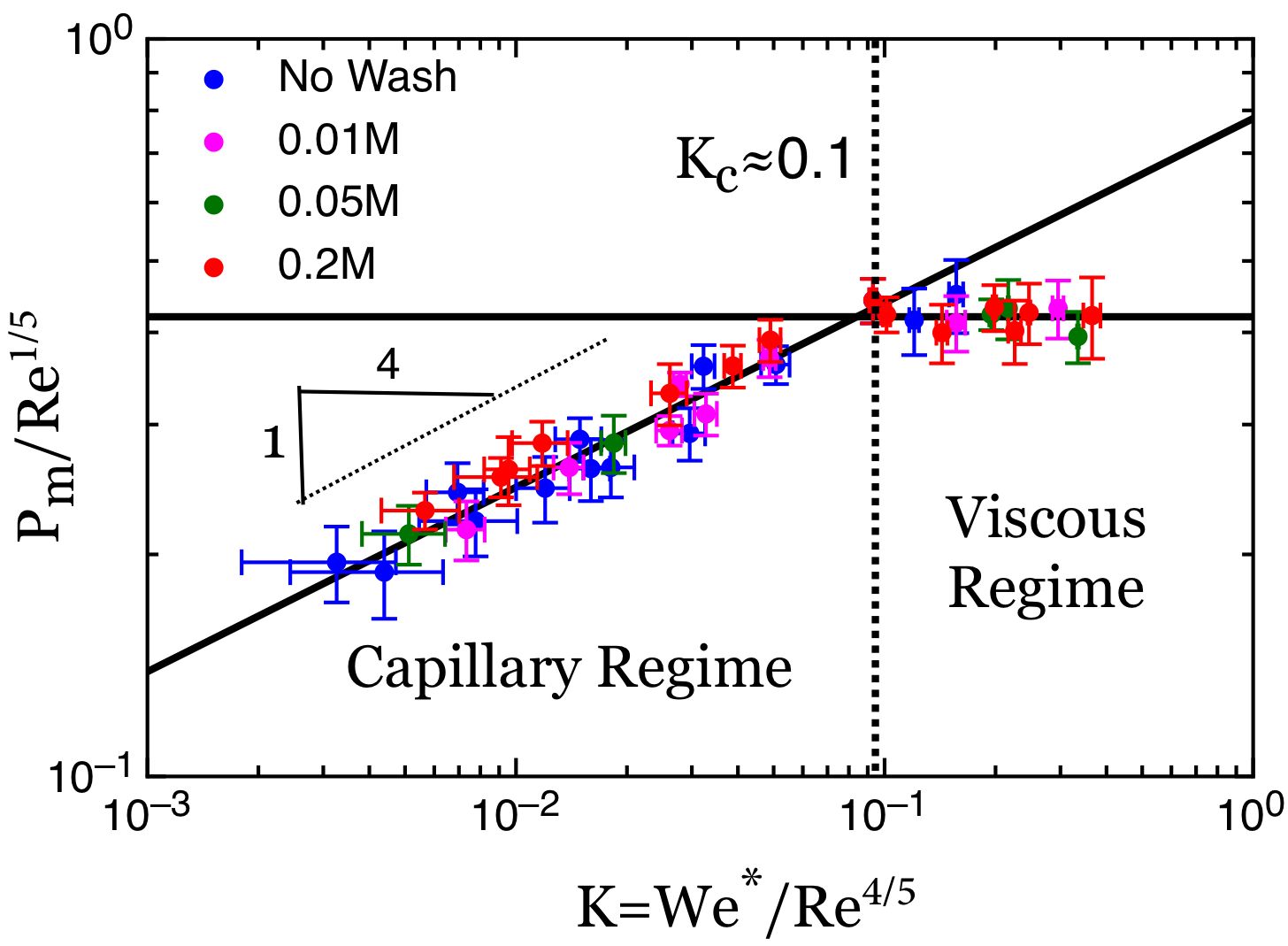}
\end{center}
\caption{Capillary-to-viscous transition for the impact behavior of eGaIn drops pre-washed in different acid concentrations. The dimensionless parameters  $P_m/Re^{1/5}$ and $K=We^{\star}/Re^{4/5}$ are used to collapse all data.}
\end{figure}

From the scaling behavior discussed in the last two sections, we can assemble a phase diagram that delineates the transition from capillary to viscous regimes. To this end we utilized the renormalized spreading factor $P_m/Re^{1/5}$ and the impact number  $K=We^{\star}/Re^{4/5}$.$^{[16]}$ Plotted log-log in terms of these variables, in the capillary regime, where $P_m \sim We^{\star 1/4}$, the data lie on a line of slope $1/4$, i.e., $P_m/Re^{1/5} \sim K^{1/4}$. In the viscous regime, where $P_m \sim Re^{1/5}$, the data lie on a horizontal line, i.e., $P_m/Re^{1/5} = const$.

As seen from Fig. 9, this way of plotting produces excellent collapse of all our data across the full range of acid concentrations, drop heights and nozzle diameters.  It also highlights the transition from capillary to viscous regime at a critical impact parameter value  $K_c\simeq 0.1$.  The observed  $K_c$ value is smaller than in Newtonian fluids, such as oil and water, for which  ${K_c}\sim 1$.$^{[16]}$ This may be attributed to the factors of the surface oxidation or drop shapes (Sec. IV), which are not included in the scaling arguments of ordinary fluids. 

\section{Conclusions}
 In this paper, we experimentally investigated the influence of surface oxidation on the impact dynamics of eGaIn liquid metal drops. We showed that the large yield stress in eGaIn exposed to air is associated with an oxide skin that can be tuned, as well as eliminated, by immersing the sample into acid solutions of varying concentration (Fig. 2). In the presence of the skin, the shear modulus $G^{\prime}$ of liquid eGaIn displays a sample size dependence (Fig. 3a), which indicates that  $G^{\prime}$ is not an appropriate intrinsic parameter. Instead, a rescaled surface modulus $G^{\prime}_s=(G^{\prime}\cdot r)/4$ characterizes the elastic response (Fig. 3b). 
 
 Skin-induced stresses also affect drop impact and spreading. In the capillary regime, where the surface energy dominates the dynamics, the oxidation level determines the impact outcome dramatically. At low Weber number, unoxidized liquid eGaIn drop can rebound while oxidized eGaIn drops always stay on the surface (Figs. 5 and 6). Rebound is enabled by the fact that chemical reduction of the oxide layer can turn the metal surface hydrophobic. 
 
 The skin effect directly affects the geometric deformation during the impact and the subsequent lateral spreading. For skin-covered liquid eGaIn in the capillary regime, the spreading factor $P_m$ does not follow the Weber number scaling $P_m\sim We^{1/4}$ observed for ordinary liquids (Fig. 7a).  We attribute this to the skin applying an extra line tension $\tau_s$ at the liquidÐair interface, resulting in an effective surface tension that combines contributions from the native material and the oxide skin. We show that this can be captured by an effective Weber number $We^{\star}=We/(1+\tau_y l/(4\sigma))$  which collapses all spreading data in the capillary regime when plotted as $P_m\sim We^{\star 1/4}$ (Fig. 7b). In particular, from the definition of  $We^{\star}$ we see that the influence of the skin increases with drop diameter $l$ and that $P_m\sim We^{\star 1/4}$ gives a condition for the effect to become significant.
 
 Viscous dissipation becomes a dominating factor only at large impact velocities. In this regime, we found no difference in spreading factor between pure and oxidized eGaIn (Fig. 8) and the usual Reynolds number scaling $P_m\sim Re^{1/5}$ is still applicable. Therefore, the resistance to spreading is mostly due to viscous drag in the bulk of the material and has very little to do with the skin.  Finally, when the renormalized spreading factor $P_m/Re^{1/5}$ is plotted as a function of impact number $K=We^{\star}/Re^{4/5}$ all data obtained under different oxidation conditions, different nozzle diameters and different impact velocities, collapses onto a single graph, with a transition between capillary and viscous regimes at  $K=We^{\star}/Re^{4/5}\simeq 0.1$.
 
 Nevertheless, the importance of the skin-induced surface stress is usually neglected in modeling the impact dynamics of molten metal drops$^{[14],[15],[18]}$. In these models, the maximum spreading factor $P_m$ was directly obtained from the balance between kinetic, native surface energy and viscous dissipation. However, it turned to be difficult to predict the spreading radius, especially in the low Weber number regime, without the loss of generality$^{[14],[15,][18]}$.Our findings suggest the cause of the difficulties is neglecting the energy stored in the oxide skin. Taking its effect into account through an appropriately rescaled Weber number $We^{\star}$ gives excellent predictability for $P_m$ (Fig. 9). 
 
 Finally, our results may also provide new insights about the impact behavior of general yield stress fluids, including dense suspension and polymer gels$^{[26], [27]}$. For these classic yield stress fluids, surface properties remain similar to ordinary liquids while the kinematic viscosity usually displays unconventional character. Therefore, their unique properties should be observed mostly in the viscous regime, which is opposite to the impact of oxidized liquid metals. 
 
 We thank Wendy Zhang and Michelle Driscoll for many useful discussions, and Qiti Guo for the help of preparing the experimental samples. This work was supported by the National Science Foundation (NSF) MRSEC program under DMR-0820054.

\end{spacing}
\end{document}